\documentclass[twocolumn,preprintnumbers,pra,showpacs]{revtex4}
\usepackage{graphicx}
\usepackage{hyperref}
\usepackage{times}

\begin{document}

\preprint{UCB-PTH-09/12, IPMU09-0048}
\title{Randall-Sundrum graviton spin determination using azimuthal angular dependence}

\author{Hitoshi Murayama$^{1,2,3}$ and Vikram Rentala$^{1,3}$}
\affiliation{$^1$ Department of Physics, University of California,
                Berkeley, CA 94720, USA}
\affiliation{$^2$ Theoretical Physics Group, Lawrence Berkeley National Laboratory,
                Berkeley, CA 94720, USA}
\affiliation{$^3$ Institute for the Physics and Mathematics of the Universe, University of Tokyo, 5-1-5 Kashiwa-no-ha, Kashiwa, Japan 277-8568}                
\date{\today}

\begin{abstract}
Quantum interference of helicity amplitudes provides a powerful tool for measuring the spins of new particles. By looking at the azimuthal angular dependence of the differential cross-section in the production followed by decay of a new particle species one can determine its spin by looking at the various cosine modes. The heavy spin-2 Kaluza-Klein (KK) graviton provides a unique signature with a $\cos{(4 \phi)}$ mode. We study the feasibility of this approach to measuring the spin of the KK graviton in the Randall-Sundrum Model at the LHC.

\end{abstract}
\maketitle

\section{Introduction \label{sec:intro}}
The Large Hadron Collider (LHC) at CERN is expected to produce a wealth of discoveries by probing the TeV scale for the first time. Apart from finally accessing the electroweak symmetry breaking scale and thus potentially discovering the elusive Higgs boson, we expect to see new physics that resolves the hierarchy/naturalness problem \cite{Weinberg:1975gm,Weinberg:1979bn,Susskind:1978ms,'tHooft:1980xb} and perhaps provides an insight into the nature of dark matter. One exciting possible solution to the hierarchy problem is the existence of warped extra dimensions \cite{Randall:1999ee,Randall:1999vf} which allows for TeV scale gravitational interactions. There are many variations of the basic theory \cite{egs} but one common feature that they share is the existence of heavy Kaluza-Klein (KK) gravitons.

For the purpose of this paper we will consider a Randall-Sundrum model with 3+1 dimensional spacetime with one additional warped extra dimension (RS1). The Standard Model fields are confined to a 3+1 dimensional TeV brane and the graviton propagates freely in the 4+1 dimensional bulk. Quantization of the graviton wave function in the extra dimension, with boundaries between the TeV brane and a Planck brane, leads to various modes which appear as heavy spin-2 fields in the 3+1 dimensional effective theory on the TeV brane. 

One of the challenges at the LHC will be to determine the spins of newly discovered particles in order to distinguish various theoretical models. The KK graviton provides a unique signature of gravitational physics at the TeV scale by virtue of its spin-2 nature. Thus, it becomes crucial to have techniques to identify its spin. 
So far, the technique proposed to measure KK graviton spin at the LHC relies on resonant graviton production followed by decay into a lepton pair \cite{Davoudiasl:1999jd,Davoudiasl:2000wi,Allanach:2000nr,Osland:2008sy,Osland:2009xe}. By looking at the polar angular dependence of the leptons relative to the beam axis, one sees a quartic behaviour of the differential cross-section.
\begin{equation}
\frac{d\sigma}{d\cos{\theta}} = A \cos^4{\theta} +  B \cos^2{\theta} +C 
\end{equation}

Recently, a new technique for measuring spin has been proposed. One can look at quantum interference of helicity states in the azimuthal angular dependence of particle decays to study their spin in a model independent way \cite{MurayamaTalk,Buckley:2007th}. The goal of this paper is to apply this technique to study the KK graviton spin and look at its feasibility at the LHC.

\section{Model Parameters}
The interaction between the massive KK gravitons and the Standard Model fields in the 4-d effective theory is given by the Lagrangian \cite{Giudice:1998ck,Han:1998sg},
\begin{equation}
{\cal L}_{int} = -\frac{1}{\Lambda} \sum_{n} {G^{(n) \mu \nu} {\cal T}_{\mu \nu}}
\end{equation}
Here, $G^{(n) \mu \nu}$ represents the $n$th KK graviton mode. ${\cal T}_{\mu \nu}$ is the stress-energy tensor of the Standard Model Lagrangian given by,
\begin{equation}
{\cal T}_{\mu \nu} = \left.  -\eta_{\mu \nu} {\cal L}_{SM} + 2 \frac{\delta {\cal L}_{SM}}{\delta g_{\mu \nu}} \right |_{ g_{\mu \nu}=\eta_{\mu \nu}  }
\end{equation}
$\Lambda$ is the coupling given by,
\begin{equation}
\Lambda = e^{-k r_c \pi} \bar{M}_{pl}
\end{equation}
where $k$ is of the order of the Planck scale, $r_c$ is the compactification radius of the extra dimension and $\bar{M}_{pl} \equiv M_{pl}/\sqrt{8 \pi}$ is the reduced Planck scale. Note the absence of KK-parity which allows the heavy graviton modes to decay into purely Standard Model particles.

The mass of the $n$th KK-graviton is given by,
\begin{equation}
 m_n = x_n \Lambda \frac{k}{\bar{M}_{pl}}
\end{equation}
where $x_n$ are the $n$th zeros of the $J_1$ Bessel function.
While studying the properties of the $n$ = $1$ KK graviton we can thus regard this theory as being dependent on only two parameters $\Lambda$ and $k$ or equivalently the dimensionless coupling $c \equiv \frac{k}{\bar{M}_{pl}}$ and $m_1$, the mass of the KK graviton of interest. 

\begin{figure}
\centering
\includegraphics[width=\columnwidth]{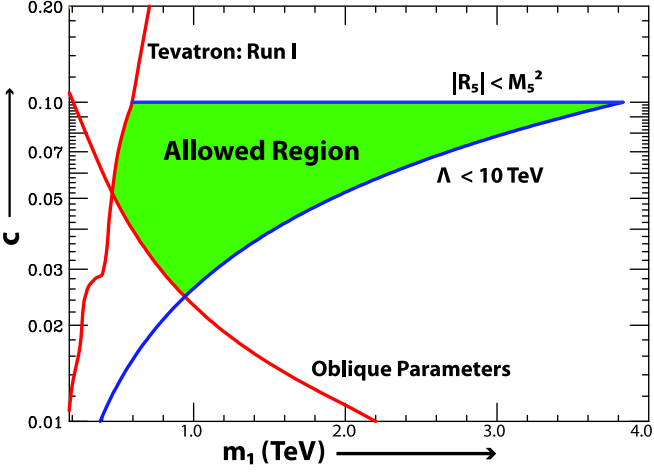}
\caption{Experimental and theoretical constraints on the KK graviton parameters in the $c-m_1$ plane. Red curves show experimental constraints and blue curves show theoretical constraints. The green shaded region shows the allowed parameter space. \label{fig:constraints}}
\end{figure}

Naturalness constraints require $\Lambda$ less than about 10 TeV. In order for an effective field theory description of gravity to be valid we require that the 5-d curvature bound, $|R_5| < {M}^2_5$, is satisfied, where $M_5$ is the 5-d Planck scale. By looking at the various theoretical and experimental constraints on the model parameters \cite{Davoudiasl:2000wi} (Figure~\ref{fig:constraints}) we expect $c$ to lie roughly between 0.01 (weakly coupled) and 0.1 (strongly coupled). We consider $m_1$ in the range of 750 GeV - 2 TeV. The decay width of the graviton to Standard Model particles can be evaluated by using the expressions given in \cite{Han:1998sg,Allanach:2000nr,Agashe:2007zd}. In the limit that decay particle masses can be neglected the decay width of the graviton is given by 
\begin{equation}
\Gamma_n = \alpha m_n{(x_n c)}^2
\end{equation}
where $\alpha$ is a constant depending upon the number of open decay channels.
If one assumes decay to only Standard Model particles the ratio ${\Gamma_1}:{m_1}$ is found to be $1.37\%$ for $c = 0.1$ (Assuming a Higgs mass of 120 GeV and decay into Standard Model particles only). This value is in disagreement with the value $1.43\%$ cited in the literature \cite{Belotelov:2006bh}.    

\section{Using Azimuthal Angular Dependence to Measure Spin}
To determine the spin of a particle $X$, we consider the production process $A+B$ $\rightarrow$ $X+Y$ where $X$ further decays to $M+N$. 
Here, $A$ and $B$ refer to beam particles or partons, $X$ is the parent particle whose spin we wish to measure. $M$ and $N$ are the daughter particles that $X$ decays into.

\begin{figure}[t]
\centering
\includegraphics[width=\columnwidth]{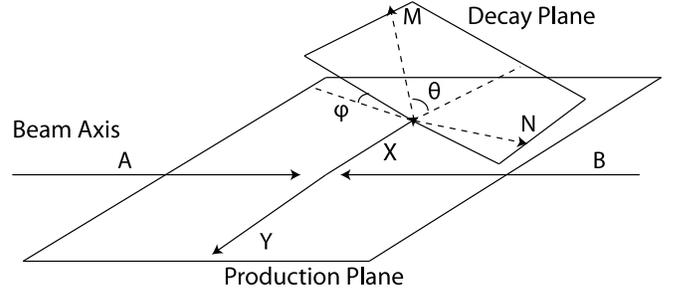}
\caption{Production and decay planes of the process $A+B$ $\rightarrow$ $X+Y$ $\rightarrow$ $M + N$. The angle $\phi$ is defined as the azimuthal angle between $\vec{p}_X$ and $\vec{p}_M$ or equivalently the angle between the production and decay planes. \label{fig:prodplane}}
\end{figure}

This gives us two planes to consider, namely the production plane (defined by the beam direction and the parent momentum) and the decay plane (defined by the parent momentum and either daughter) (Figure~\ref{fig:prodplane}).

Now consider the daughter $M$ with momentum $\vec{p}_M$. The angle it makes with the parent momentum $\vec{p}_X$ is defined to be $\theta$. Projecting out the component of $\vec{p}_M$ parallel to $\vec{p}_X$ and looking at the angle between the residual vector and the production plane we define an angle $\phi$. Thus, $\phi$ describes azimuthal rotations of the vector $\vec{p}_M$ in the $x-y$ plane with $\vec{p}_X$ taken to be the $z$-axis. From the figure it is clear that, equivalently $\phi$ can be defined as the angle between the production plane and the decay plane. More explicitly, we define the two vectors,
\begin{equation}
\vec{p}_{prod} = \vec{p}_A \times \vec{p}_X 
\end{equation}
and,
\begin{equation}
\vec{p}_{decay} = \vec{p}_X \times \vec{p}_M
\end{equation}
Then,
\begin{equation}
\cos{\phi}  = \hat{p}_{prod} \times \hat{p}_{decay} 
\end{equation}
Here $\hat{p}$ denotes the normalized vectors.

In the limit of the narrow width approximation, the amplitude can be split into ${\cal M}_{prod}$ and ${\cal M}_{decay}$. 
\begin{equation} 
{\cal M}_{prod} = \langle X,Y| {\cal T}_{prod} |A,B \rangle
\end{equation}
\begin{equation}
{\cal M}_{decay}(\phi) = \langle M,N, \phi  | {\cal T}_{decay}|X \rangle
\end{equation}
where we have explicitly shown the $\phi$ dependence of the the final state and decay amplitude. 
We also have,
\begin{equation}
{\cal M}_{decay}(\phi) = \langle M, N (\phi = 0) | e^{+iJ_z \phi} {\cal T}_{decay} | X \rangle
\end{equation}
where $J_z$ generates rotations about the $\vec{p}_X$ direction. We can now think of the rotation operator as acting on the interaction ${\cal T}$-matrix plus ket, rather than on the bra. Assuming, ${\cal T}_{decay}$ is rotationally invariant, we only need to consider rotations of the particle $X$ about its own momentum axis. In this case,
\begin{equation}
 J_z = \vec{J} \cdot \hat{p} = (\vec{s} + \vec{r} \times \vec{p}) \cdot \hat{p} = \vec{s} \cdot \hat{p} = h
\end{equation}

Thus, rotations about the momentum axis of a given helicity state, $h$ for $X$ only produce a phase $e^{+i h \phi}$. So, 
\begin{equation}
{\cal M}_{decay}(\phi) = e^{+i h \phi} {\cal M}_{decay}(\phi =0)
\end{equation}
Thus, allowing for production over all possible helicity states of $X$ we must sum coherently over all possible amplitudes, and so, the differential cross section takes the form
\begin{equation}
\frac{d\sigma}{d\phi} \propto \left | \sum_{h} {\cal M}_{prod} e^{+i h \phi} {\cal M}_{decay}(\phi =0) \right |^2
\end{equation}
Here, $h$ runs from $-s$ to $+s$ where $X$ has spin $s$. From this it is clear that, if we look at the differential distribution $d\sigma/d\phi$, interference between various helicity states is responsible for a non-trivial $\phi$ dependence,
\begin{equation}
\frac{d\sigma}{d\phi} = A_0 + A_1 \cos{(\phi)} + A_2 \cos{(2 \phi)} + ....+ A_{2s} \cos{(2s \phi)}.
\end{equation}
Note the absence of $\sin{(n \phi)}$ modes, which would be present in the case of CP violating processes.

The Standard Model has no particles with spin greater than 1 and so the largest mode from the Standard Model would only be $\cos{(2 \phi)}$, corresponding to $X$ being a gauge boson. We can now see the unique signature that the KK graviton will produce, namely a $\cos{(4 \phi)}$ mode. 

Also, we note that this result is valid in any reference frame but the size of the coefficients $A_i$ will be different in different reference frames. To maximize this unique signature for the KK graviton, we need to choose a reference frame in which $A_4/A_0$ has a large value. 

\section{Signal and Background}
We assume that the mass of the graviton will be well measured using resonant graviton production through the process $pp \rightarrow G \rightarrow l^+ l^-$ \cite{Davoudiasl:1999jd,Davoudiasl:2000wi}.

The process we are considering is $pp$ $\rightarrow$ $G$ + jet followed by $G$ $\rightarrow$ $l^+$ $l^-$ where $l$ are muons or electrons. 
The dominant parton level subprocess comes from $gg$ $\rightarrow$ $Gg$ with subdominant $q$($\bar{q}$) $g$ $\rightarrow$ $G$ $q$$(\bar{q})$ and the crossed channel $q$ $\bar{q}$ $\rightarrow$ $G$ $g$. Here $G$ represents the graviton, $g$ represents gluons and $q$ represents the various quarks.

The Standard Model background comes from the subdominant channels with $G$ replaced by an off-shell $Z$, $\gamma$. This is the exact analog of Drell-Yan background in resonant graviton production. Cutting on the invariant mass of the lepton pair in a mass window around the graviton mass gets rid of most of the background. The Standard Model background consists of spin-1 states and can not give any contribution to $A_4$. At most it can affect the value of $A_0$ and dilute the value of $A_4/A_0$.

\section{Calculating the differential cross-section}

\subsection{Zero Rapidity Frame \label{sec:calc}} 
\begin{figure}
\centering
\includegraphics[width=\columnwidth]{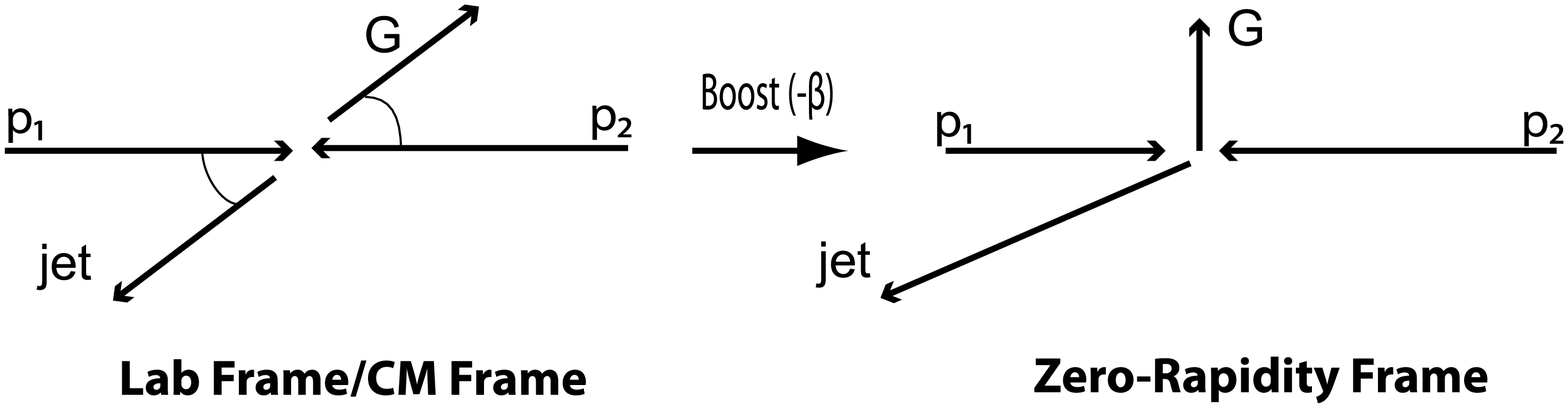}
\caption{Boost from center of mass or laboratory frame to the Zero Rapidity frame \label{fig:ZRframe}}
\end{figure}

The dilepton + jet events that we are looking for are fully reconstructible at the LHC. The key reason for this is that we have a signature with no missing energy-momentum which in turn is a direct consequence of the absence of KK parity. The graviton 4-momenta should have minor errors compared to the jet reconstruction since it is reconstructed from the di-lepton 4-momenta. As previously mentioned the size of the non-zero coefficients $A_i$ are frame dependent and so we must choose a frame in which the normalized coefficient $S_4 \equiv |A_4/A_0|$ is large. It was found that in the center of mass frame of the partonic processes, $S_4$ was larger than in the lab frame. However, transforming from the lab-frame to the center of mass frame would have an error dependent on the error of the jet reconstruction. To avoid this error and still make an improvement in the signal, we studied $S_4$ in the zero rapidity frame of the graviton, i.e. the frame where the graviton is purely transverse to the beam axis (Figure~\ref{fig:ZRframe}). The reason for this is the boost factor can be calculated from just the graviton momentum in the lab frame which is well reconstructed from the leptons.

\subsection{Cuts}
The first set of cuts used included a pseudo-rapidity ($|\eta|<2.5$) cut and $p_T > 20$ GeV cut for the jet. The second set of cuts was a mass-window cut on the invariant mass distribution of the lepton pair. This gets rid of a large portion of the Standard Model background. The size of the window was determined by detector resolution at ATLAS \cite{:1999fq,:1999fr} for an $e^+ e^-$ pair. The third set of cuts involved rapidity cuts ($|\eta|<2.6$) on each of the leptons with a requirement that $p_T$ $>$ 10 GeV for either one of the leptons and $p_T$ $>$ 20 GeV for the other. An isolation cut, $\Delta r \equiv \sqrt{(\Delta \eta)^2 + {\Delta \phi}^2}$ $>$ 0.7, was imposed between the lepton and the jet. However, the third set of cuts affects the angular distribution of the leptons and can create false cosine modes in the differential distribution. 

\begin{figure}
\centering
\includegraphics[width=\columnwidth]{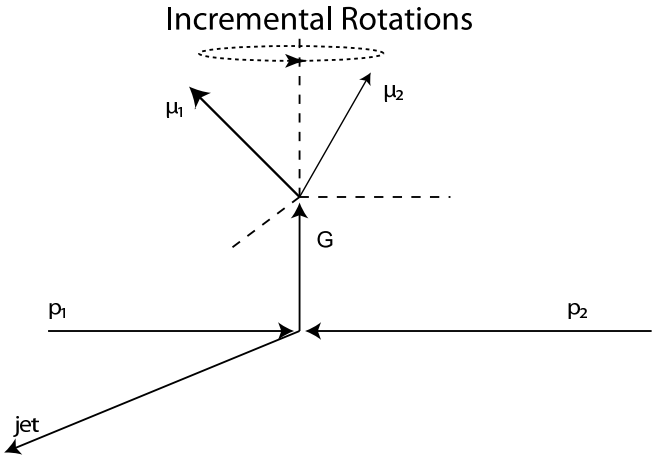}
\caption{The leptons are rotated about the graviton momentum axis in the zero rapidity frame. The dilepton + jet momenta must be reboosted to the lab frame at each step to make sure that they pass the cuts.\label{fig:rotinZRframe}}
\end{figure}

To solve this problem one imposes `rotationally invariant cuts', first introduced in \cite{Buckley:2008pp}. Thus, it is not sufficient for the observed lepton to simply pass these cuts, the leptons are rotated around the graviton momentum axis in small increments and at each step it is checked that the lepton passes the cuts. The added complication is that the rotations must be made in the zero-rapidity frame to preserve rotational invariance in that frame (Figure~\ref{fig:rotinZRframe}).
So, the procedure is as follows:
\begin{enumerate}
\item First, reconstruct the event completely using the dilepton and jet signals.
\item Calculate the boost factor to take us from the lab frame to the zero-rapidity frame of the graviton. 
\item Boost all momenta to the zero-rapidity frame. Rotate the leptons about the graviton momentum direction by a small angle, say $1^{\circ}$. 
\item Reboost the new lepton and jet momenta to the lab frame. Check if they pass the cuts, if they don't throw out the event. 
\item If they do pass the cuts go back to step 3.
\item Repeat this procedure until we have made a full $360^{\circ}$ rotation of the lepton momenta about the graviton momentum axis in the zero rapidity frame. 
\end{enumerate}
This procedure ensures that the cuts do not affect the azimuthal angular distribution in the zero-rapidity frame. 

\subsection{Simulations}
We used HELAS \cite{Murayama:1992gi} with spin-2 particles \cite{Hagiwara:2008jb} to calculate the helicity amplitudes for the graviton scattering process. LHApdf \cite{Whalley:2005nh} was used to fold in the parton distribution functions for the protons. We used the pdf set CTEQ6L \cite{Pumplin:2002vw}. An adaptive Monte-Carlo package, BASES \cite{Kawabata:1985yt}, was used to perform the integration over phase space and produce the differential cross-section $d\sigma/d\phi$. 

\section{Determining the coefficients of the various Cosine modes}
Once we have the binned distribution $d\sigma/d\phi$ with $2n$ bins (For the purposes of calculation in this paper we used 50 bins), we try to fit coefficients of the form 
\begin{eqnarray}
  x_i & \equiv & \frac{1}{\mathrm{Binsize}}\displaystyle\int^{\frac{2 \pi i}{2n}}_{\frac{2 \pi(i-1)}{2n}} \frac{d\sigma}{d\phi} d\phi \nonumber \\
&=& \frac{1}{2\pi / 2 n}\displaystyle\int^{\frac{2 \pi i}{2n}}_{\frac{2 \pi(i-1)}{2n}}  \left[\sum_{j=0}^{n-1} A_j \cos{(j\phi)}\right. \nonumber \\
& & \left. \qquad \qquad + \sum_{j=1}^{n} B_j \sin{(j\phi)} \right] d\phi
\end{eqnarray}
where $i$ runs over 0, 1, 2, ..., 2$n-$1. The integration accounts for the binning process and the $2n$ coefficients $A_0$, ..., $A_{n-1}$, $B_1$, ...., $B_n$ correspond to the strengths of the various cosine and sine modes that can be resolved from each other. 

Thus, we have a simple linear relationship between the $2n$ binned values of $d\sigma/d\phi$ ($x_i$) and the $2n$ binned-Fourier coefficients ($y_j$) of the form $x_i = p_{ij} y_j$. Here, $p_{ij}$ are either of the form $\displaystyle\int^{\frac{2 \pi i}{2n}}_{\frac{2 \pi(i-1)}{2n}} \cos{(j\phi)} d\phi$ or  $\displaystyle\int^{\frac{2 \pi i}{2n}}_{\frac{2 \pi(i-1)}{2n}} \sin{(j\phi)} d\phi$.

Now, we can simply invert this matrix for a given value of $n$ to recover the amplitudes of the various harmonics. For the $d\sigma/d\phi$ distribution for the graviton we expect to see only the coefficients $A_0$, ..., $A_4$ to be non-zero. Also, since the beams are identical, we expect to see only the even cosine modes. The odd cosine modes drop out since they flip sign when the beams are switched ($\phi \rightarrow \pi - \phi$).

\section{Results and Discussion}
\begin{figure}
\centering
\includegraphics[width=\columnwidth]{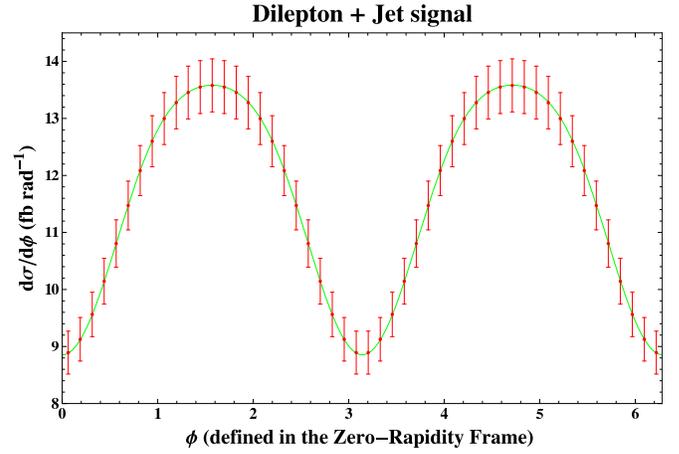}
\caption{Differential distribution ($\frac{d\sigma}{d\phi}$) for $m_1$ = 1 TeV and c = 0.05. A strong $\cos{(2\phi)}$ mode can be seen but there is also a $\cos{(4\phi)}$ component. The theoretical curve (produced from simulations) is shown in green. The red dots indicate the binned values, with error bars corresponding to Gaussian errors for a luminosity of 500 fb$^{-1}$\label{fig:dsigdphi}}
\end{figure}

\begin{figure}
\centering
\includegraphics[width=\columnwidth]{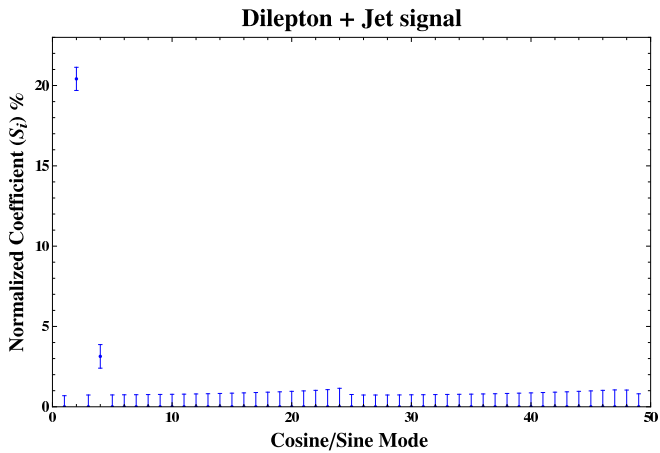}
\caption{Fitted cosine coefficients of the binned differential cross-section shown in Figure~\ref{fig:dsigdphi} corresponding to 50 bins. The first 25 modes label the normalized cosine modes, the next 25 show the sine modes. The large 0-mode which would be $100\%$ is not shown). See text for how the error bars in this plot are calculated using error bars from the binned differential cross-section. \label{fig:FT}}
\end{figure}

Simulations were done for the process $pp \rightarrow$ $e^+$$e^-$$j$ at 7 TeV beam energy using a dilepton invariant mass window cut around the graviton mass. Figure~\ref{fig:dsigdphi} shows the $d\sigma/d\phi$ distribution for a 1 TeV graviton with $c = 0.05$. Figure~\ref{fig:FT} shows the normalized fitted coefficients. The normalized cosine coefficients ($S_i$) are shown in the first 25 bins, with the zero mode suppressed. The next 25 bins show the sine modes. The size of the $S_4$ coefficient is 3.14\%. Note the absence of odd cosine modes, this arises from the fact that we are using identical beams.

To look at the dependence of the signal on graviton mass, simulations were performed for $c = 0.1$ and $m_1$ = 750 GeV, 1 TeV, 1.5 TeV and 2 TeV. The results are summarized in Table~\ref{tab:mvar}. 
\begin{table}[h]
\centering
\begin{tabular}{|c|c|c|c|c|c|}
\hline
 $m_1$ (TeV) & $\Delta m$ (GeV) &  $\sigma_{total}$ (fb)& $\sigma_{bgd}$ (fb)   & $S_2$ & $S_4$  \\ \hline \hline
 $0.75$   & $24.4$ &  $871.7$ & $0.39$   &  $20.00\%$ &  $3.50\% $\\ \hline
 $1.0$    & $30.7$&  $229.8$ & $0.15$  &  $20.48\%$ &  $3.16\% $\\ \hline
 $1.5$  & $42.8$&  $28.7$ & $0.03$  & $20.70\%$  & $1.52\%$ \\ \hline
 $2.0$   & $55.0$ &  $5.52$ & $0.01$  & $20.08$\% &  $0.80\%$\\ \hline
\end{tabular}
\caption{Signal strength $S_4 \equiv |A_4/A_0|$ as a function of the mass of the graviton. $c$ = 0.1 for all entries. $S_2$ is shown for comparison. The mass window (based on the ATLAS detector resolution for $e^+e^-$ invariant mass \cite{:1999fq,:1999fr}) cuts out most of the background. \label{tab:mvar}}
\end{table}
The total cross-section decreases rapidly with graviton mass as expected. The background is negligible and, as we will see in the next paragraph, has little effect even if the coupling $c$ is reduced. The main concern is therefore the decrease in $S_4$ and the low cross-section at large values of $m_1$.

The results for a 1 TeV graviton at different values of the coupling $c$ are shown in Table~\ref{tab:cvar}. In the absence of cuts the graviton cross-section is expected to approximately scale like $c^2$. The Standard Model background level is 0.15 fb which is $\sim 5\%$ at $c$ = 0.01. The value of $S_4$ is expected to be diluted slightly by the background because of a corresponding $5\%$ increase in $A_0$. As $c$ is increased, the background as a percentage of the cross-section decreases and $S_4$ is restored to its maximum strength.

\begin{table}[h]
\centering
\begin{tabular}{|c|c|c|c|}
\hline
 $c$   & $\sigma_{total}$ (fb)& $S_2$ & $S_4$  \\ \hline \hline
 $0.01$    &     $3.27$& $18.62\%$& $3.05\%$ \\ \hline
 $0.02$   &     $12.51$ & $20.02\%$& $3.15\%$ \\ \hline
 $0.05$    &     $72.75$& $20.42\%$& $3.14\%$ \\ \hline
 $0.1$   &     $229.8$& $20.48\%$ & $3.16\%$  \\ \hline
\end{tabular} 
\caption{Signal strength $S_4 \equiv |A_4/A_0|$ as a function of the coupling c. All entries are for $m_1$ = 1 TeV. $S_2$ is shown for comparison. The Standard Model background cross-section is 0.15 fb.\label{tab:cvar}}
\end{table}

\section{Error Analysis}
As we have seen, the effect of background is small and does not contribute to $A_4$. Its only effect is to dilute the normalized coefficient $S_4$. Thus, the experimental error will be determined by event statistics. We assumed Gaussian errors ($\Delta x_j = x_j \frac{\sqrt{N_j}}{N_j}$) in the $j$th bin assuming $N_j = {\cal L} {\sigma} \frac{x_j}{\sum x_j}$ events in each bin for integrated luminosities ${\cal L}$ of 10, 100 and 500 fb$^{-1}$. Since, the coefficients $A_i$ are determined from the binned values $x_j$ through a simple linear relationship (via the matrix $q_{ij} = p^{-1}_{ij}$). It is then straightforward to work out the errors in the normalized coefficients ($\Delta S_i$).
\begin{equation}
\Delta S_i = \sqrt{\sum_j \left(\frac{q_{ij}}{A_0} - \frac{S_i}{A_0}q_{0j}\right)^2 \Delta x_j^2}
\end{equation}
The first term in the paranthesis arises from the simple linear relationship between $A_i$ and $x_j$. The second term comes from the error associated with the normalization factor $A_0$. The relative errors ($\frac{\Delta S_4}{S_4}$) for various integrated luminosities at different points in the parameter space of the model are given in Table~\ref{tab:deltas4}. A value $>1$ for the relative error indicates that statistics would be poor and give no reason to doubt $S_4$ being consistent with 0. A value of $0.20$ or less indicates at least a $5\sigma$ effect indicating high likelihood of confirmation of the spin-2 nature of the KK graviton. 
\begin{table}[h]
\centering
\begin{tabular}{|c|c|c|c|c|}
\hline
 $m_1$ (TeV) & $c$   & 10 fb$^{-1}$ & 100 fb$^{-1}$& 500 fb$^{-1}$ \\ \hline \hline
 $0.75$ &$0.1$     & $0.43$ & $0.14$& $0.06$\\ \hline
 $1.0$  & $0.01$   &  $8.03$  & $2.54$ & $1.14$\\ \hline
 $1.0$  & $0.02$    & $3.97$  & $1.26$& $0.56 $\\ \hline
 $1.0$  & $0.05$    &  $1.65$  & $0.52$& $0.23$ \\ \hline
 $1.0$  & $0.1$     & $0.93$ & $0.29$& $0.13$\\ \hline
 $1.5$ &$0.1$     & $5.42$ & $1.71$& $0.77$\\ \hline
 $2.0$  &$0.1$     & $23.52$ & $7.44$& $3.32$\\ \hline
\end{tabular} 
\caption{Statistical Error $\Delta S_4 / S_4$ for different integrated luminosities for the process $pp \rightarrow e^+ e^- j$. $\Delta S_4 / S_4 < 0.5 (0.71)$ corresponds to a $2\sigma$ confirmation of the graviton spin, and $\Delta S_4 / S_4 < 0.2 (0.28)$ corresponds to a $5\sigma$ confirmation. The values in brackets denote the $2\sigma$ and $5\sigma$ confidence levels if one includes $\mu^+ \mu^- j$ production channels as well.\label{tab:deltas4}}
\end{table}

Alternatively, if one requires only a 95$\%$ confidence level ($2\sigma$) effect then a value of $0.5$ or less for $\Delta S_4 / S_4$ should suffice. If we additionally assume information from $\mu^+ \mu^- j$ statistics in addition to the $e^+e^- j$ channel (assuming that detector resolution for the invariant mass is the same for both lepton species), then we can see a factor 2 improvement in the statistics. This would in turn result in a factor $\sqrt{2}$ drop in the error. Thus, in this case the parameter space in Table~\ref{tab:deltas4} for which $\Delta S_4 / S_4 < 0.71$ would correspond to potential for a $2\sigma$ confirmation of the graviton spin, and $\Delta S_4 / S_4 < 0.28$ would correspond to a $5\sigma$ confirmation.

\section{Comparison with resonant graviton production method and distinction from spin-0}
Osland et al. \cite{Osland:2008sy, Osland:2009xe} consider the resonant graviton production process $pp \rightarrow G \rightarrow l^+l^-$ to measure the spin of the graviton using the quartic angular dependence of the polar angle of the lepton. This results in a center-edge asymmetry ($A_{CE}$) in the differential distribution $d\sigma/d\cos{\theta}$. Their results indicate ($2\sigma$) identification of the graviton spin for $c = 0.01$ and $10^{-1}$ fb of luminosity for masses upto 1.1 TeV. For $c = 0.1$ they claim identification upto masses of $2.4$ TeV.

The azimuthal angular dependence method that we considered has inherently lower statistics compared to resonant graviton production because of the extra recoiling jet. Our method suffers from lower statistics, but given higher luminosities, it can still provide an independent confirmation of the KK graviton spin for a large region of the expected parameter space of the KK graviton.

The center-edge asymmetry method can distinguish a spin-1 particle ($Z'$) from a KK graviton more readily than it can distinguish it from a spin-0 particle.

Our method proves complementary, since the KK graviton also produces a large $\cos(2\phi)$ mode ($S_2 \sim 20\%$) and can thus easily be distinguished from a scalar which would not produce any non-zero modes. The results for $\Delta S_2 / S_2$ are shown in Table~\ref{tab:deltas2}.

\begin{table}[h]
\centering
\begin{tabular}{|c|c|c|c|c|}
\hline
 $m_1$ (TeV) & $c$   & 10 fb$^{-1}$ & 100 fb$^{-1}$& 500 fb$^{-1}$ \\ \hline \hline
 $0.75$ & $0.1$     & $0.07$ & $0.02$& $0.01$\\ \hline
 $1.0$  & $0.01$   &  $1.30$  & $0.41$ & $0.18$\\ \hline
 $1.0$  & $0.02$    & $0.62$  & $0.19$ & $0.09$\\ \hline
 $1.0$  & $0.05$    &  $0.25$  & $0.08$& $0.04$ \\ \hline
 $1.0$  & $0.1$     & $0.14$ & $0.04$& $0.02$\\ \hline
 $1.5$ &$0.1$     & $0.39$ & $0.12$& $0.06$\\ \hline
 $2.0$  &$0.1$     & $0.93$ & $0.29$& $0.13$\\ \hline
\end{tabular} 
\caption{Statistical Error $\Delta S_2 / S_2$ for different integrated luminosities for the process $pp \rightarrow e^+ e^- j$. $\Delta S_2 / S_2 < 0.5 (0.71)$ corresponds to a $2\sigma$ distinction from a spin-0 particle, and $\Delta S_2 / S_2 < 0.2 (0.28)$ corresponds to a $5\sigma$ distinction. The values in brackets denote the $2\sigma$ and $5\sigma$ confidence levels if one includes $\mu^+ \mu^- j$ production channels as well.\label{tab:deltas2}}
\end{table}

Assuming, as before inclusion of $\mu^+$$\mu^-$ statistics $\Delta S_2 / S_2 < 0.71$ corresponds to a $2\sigma$ distinction from a spin-0 particle, and $\Delta S_2 / S_2 < 0.28$ corresponds to a $5\sigma$ distinction. In regions of the parameter space of $m_1$, where the Standard Model background is comparable to the cross-section of interest (Table~\ref{tab:mvar}), the confidence levels are altered slightly because the off-shell $\gamma$ and $Z$, being spin-1, contribute to the $A_2$ coefficient.

From the table, we can see that even with 10 fb$^{-1}$ of luminosity, the spin-0 hypothesis can be ruled out for a large portion of the allowed parameter space. Thus, our method proves complementary to the approach by Osland et al. by ruling out spin-0 more easily than spin-1. In both methods the distinction from spin-0 can be made from comparable integrated luminosities.

\section{Summary and Conclusion}
We studied the process $pp \rightarrow G$ jet $\rightarrow l^+ l^-$ jet and looked at the differential distribution $d\sigma/d\phi$. The distribution was found to have a $\cos{(4\phi)}$ mode, characteristic of a spin-2 particle, with strength parametrized by $S_4$. The parameter $S_4$ was $\sim 3\%$ for values of $m_1$ below a TeV. As we go to higher graviton masses the signal drops off, but what is of more concern is the drop in cross-section with large $m_1$ or low values of $c$. Both these scenarios are unlikely to occur in conjunction because of naturalness constraints (See Figure~\ref{fig:constraints}). 

In conclusion, observing higher cosine modes ($>2$) in the differential distribution would be a clear signal of beyond Standard Model physics. Observing the $\cos(4 \phi)$ mode at the LHC would be a strong indicator of gravitational physics at the TeV scale. If the coupling is strong enough $\sim$ 0.05 or greater and the mass is sufficiently low $\sim$ 1 TeV or less, we expect to have a clear signal of the spin-2 nature of the KK graviton at the LHC.

For regions of parameter space with larger masses or lower couplings, the azimuthal angular dependence of the cross-section is still useful in ruling out the spin-0 hypothesis and this can be done for fairly low luminosities $\sim$ 10 fb$^{-1}$ as well.

This method provides an important complementary and independent approach to measuring the spin of the KK graviton, as compared to the method of using polar angular dependence from resonant KK-graviton production.

\begin{acknowledgments}
 The authors would like to express their thanks to William Klemm, Kai Wang, Mihoko Nojiri and Matthew Buckley for their help and feedback. This work was supported in part by World Premier International Research Center Initiative (WPI Initiative), MEXT, Japan, in part by the U.S. DOE under Contract DE-AC03-76SF00098, and in part by the NSF under grant PHY-04-57315. 
\end{acknowledgments}

\end{document}